\documentclass{article}

\usepackage{arxiv}

\usepackage[utf8]{inputenc} 
\usepackage[T1]{fontenc}    
\usepackage{hyperref}       
\usepackage{url}            
\usepackage{booktabs}       
\usepackage{amsfonts}       
\usepackage{nicefrac}       
\usepackage{microtype}      
\usepackage{lipsum}		
\usepackage{graphicx}
\usepackage{doi}
\usepackage[table]{xcolor}
\usepackage{setspace}

\title{Design an IT Policy Implementation Plan}

\author{ \href{https://orcid.org/0000-0003-3314-4281}{\includegraphics[scale=0.06]{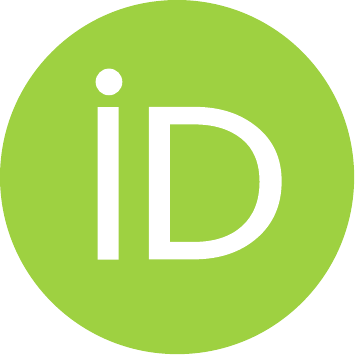}\hspace{1mm}Saman Sarraf, SMIEEE}\thanks{Corresponding author} \\
	Senior Member, IEEE\\
	School of Technology\\
	Northcentral University\\
	San Diego, CA 92037\\
	\texttt{samansarraf@ieee.org} \\
	\And
	\href{https://orcid.org/0000-0000-0000-0000}{\includegraphics[scale=0.06]{orcid.pdf}\hspace{1mm}Milton Kabia, PhD, DHA}\\
	Dissertation Chair\\
	School of Technology\\
	Northcentral University\\
	San Diego, CA 92037\\
	\texttt{mkabia@ncu.edu} \\
}

\renewcommand{\shorttitle}{\textit{IT Policy Implementation Plan}}

\begin{document}
\maketitle
\setstretch{1.25}
\begin{abstract}
	Information technology (IT) companies implement multi-dimensional policy plans that include procedures, sub-plans, and instructions to outline their business scopes, targets, and communications. This work outlined the IT policy implementation plan designed by an imaginary company with a random name called Northcentral Cloud Consulting Firm (NCCF), containing proposed IT policies, milestones and roadmaps, control framework, stakeholder responsibilities, knowledge transfer plan, and leadership roles. As NCCF’s major customers seek data-driven solutions in cloud computing, the NCCF IT policy plan provides various data policies, including security and proper usage of machine learning services. The plan offers a detailed roadmap of its financial, geographical, and reputational expansion within three years. The IT policy plan also compromises an IT risk management, contingency, and emergency communication plan, mainly for protecting data and business continuity. Stakeholder responsibilities are incorporated into the IT policy plan, as NCCF considers any engagement with its customers as a collaborative effort in which both parties have and share several responsibilities. 
\end{abstract}


\section{Introduction}
An information technology (IT) policy implementation plan is designed to provide a clear direction and instruction to internal leadership, developers, and staff who draft and build a company’s policies. Such a plan also defines stakeholders’ responsibilities in engagements with the company \cite{niemimaa2017information}. In most cases, IT companies built the strategic plans accompanied by policy implementation plans that enable them to define and develop more realistic policies and solutions \cite{garson2006public}. A standard IT policy implementation plan consists of several components such as a) assignment of responsibilities, b) scheduling, c) resource allocation, d) evaluation metrics, e) contingency plan, and f) communication plan \cite{castro2012automation}\cite{sabbaghi2017hybrid}. 
\newline
An IT policy implementation must be aligned with the company’s business needs and scopes to ensure organization growth from financial, geographical, and reputational aspects \cite{dent2015aligning}. The degree of alignment between the implementation plan and its business plan depends on its priorities and targets as determined by the leadership \cite{high2014implementing}. This work focused on an implementation plan and discussed: a) IT policies, b) milestones and roadmap, c) control framework, d) stakeholders’ responsibilities, e) communication and training plan, and leadership involvement. 
\section{Proposed IT Policies}
\label{sec:headings}
Strategy in IT is a comprehensive road map outlining high-level instructions that should be employed to satisfy requirements and targets designed by an enterprise \cite{van2004strategies}\cite{zahra1993business}. The enterprise’s strategic plan describes such targets encompassing technical and business aspects as a written document focusing on organizational growth’s most impacting factors \cite{drnevich2013information}\cite{turban2003information}\cite{sarraf2019mcadnnet}. 

\subsection{NCCF Background}
Our imaginary company so-called NCCF offers a broad range of consultation and services to customers in the United States western and central regions. Such customers often seek to migrate their infrastructure to a cloud environment that the migration process includes proof of concept (POC) and deployment phase. NCCF’s subject matter experts are split into two primary groups: a) consulting staff and b) development teams focusing on data lake and productization. The primary mission defined by NCCF refers to providing cloud-based consultation and services through reliable solutions deployable in a cloud provider-agnostic manner at the most optimal cost and performance. NCCF plans to design and implement repeatable offerings to address customers’ data problems and projects, including big data and scalable machine learning, allowing them to enhance and improve their migrated infrastructure dynamically. The secondary mission of NCCF is to concentrate on designing novel artificial intelligence services and custom machine learning modeling, allowing NCCF to overcome its competitors in the cloud market.
\subsection{Strategic Targets}
\begin{figure}[h!]
	\centering
	\includegraphics[width=0.8\linewidth]{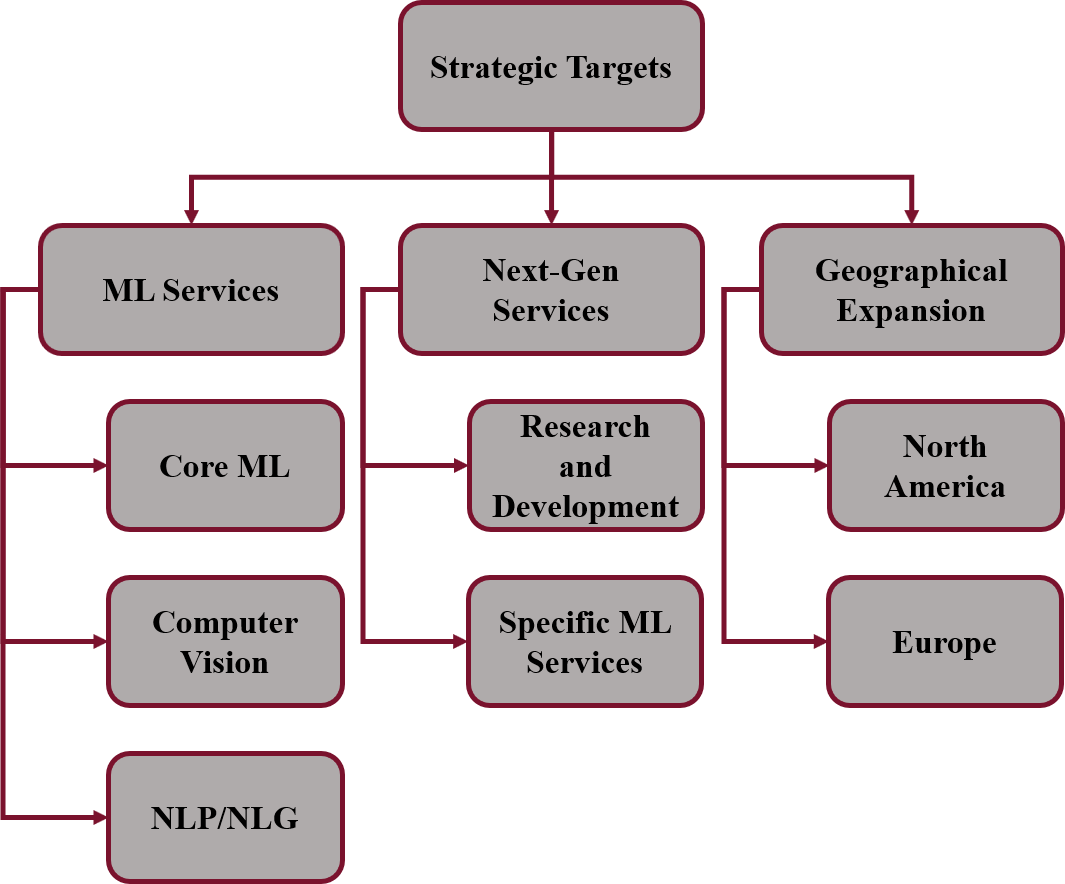}
	\caption{Strategic targets defined by NCCF}
	\label{fig1}
\end{figure}

In the light of expanding cloud services and conquering the market, NCCF has defined two primary strategic business and technical targets, including a) scalable machine learning custom modeling and repeatable offering and b) next-generation market demand cloud-based software architecture design and development. NCCF has decided to design and develop custom machine learning applications per customers’ requests and work backward from the customers’ desires to deliver projects. The current trend in machine learning technologies demonstrates the high demand for computer vision applications, natural language processing, and generation considered in the NCCF strategic plan \cite{bossmann2016top}\cite{haenlein2019brief}\cite{li2017applications}\cite{sarraf2021comprehensive}\cite{sarraf2016big}. Figure \ref{fig1} illustrates the strategic targets planned by NCCF, including ML services, next-gen services, and geographical expansion.
\subsection{IT Infrastructure}
The current IT infrastructure of NCFF is composed of a) headquarters networks, b) Internet network c) branch wide area network (WAN) connections, d) storage and database, e) cloud servers for backup and recovery, f) patch panel connectors, g) routers and switches h) customer-government connections and i) main servers hosting POC and productized applications. According to an internal policy, each headquarter includes a server room, conference room, training room, open space work areas, and managerial office spaces. Per NCCF IT policy, the servers, including main servers, production server, databases, storages, are hosted in the Amazon Web Services  cloud, and the NCCF staff and developers must host any NCCF and its customers’ data in the cloud and implement any POC or software entities using the AWS cloud services \footnote{https://aws.amazon.com/}.
\subsection{NCCF Policy and Compliance}
NCCF offers the functional and technical services previously mentioned and specific solutions defined in the strategic plan to customers with various industrial domains, including IT, healthcare, security, and finance. NCCF has established a policy and compliance portfolio to ensure that proposed solutions, including design, software development cycle, data security, and delivered software entities, meet the standard policies. The NCCF compliance portfolio illustrated in Table \ref{table1} includes a) National Institute of Standards and Technology (NIST) \cite{mahardika2017manajemen} b) Health Insurance Portability \& Accountability Act (HIPAA) \cite{mahardika2017manajemen} c) Financial Industry Regulatory Authority (FINRA) \cite{cole2007financial} d) Payment Card Industry Data Security (PCI DSS) \cite{miller2014pci} e) Control Objectives for Information and Related Technologies (COBIT) \cite{young2014cobit} f) Amazon Governance, Risk and Compliance (GRC) \cite{van2004strategies} and g) Microsoft General Data Protection Regulation (GDPR)\footnote{https://www.microsoft.com/en-us/trust-center/privacy/gdpr-overview}. The NCCF leadership added two new security compliances to enhance the IT policy plan, including the Health Information Technology for Economic and Clinical Health Act known as HITECH Act \cite{burde2011hitech} and the National Institute of Standards and Technology Cybersecurity Framework known as CSF \cite{yvon2020exploring}. Employing the HITECH Act and NIST CSF enabled the organization to expand its services into healthcare, economics, and cybersecurity domains and attract more customers. 
\begin{table}[h!]
	\caption{The objective of adopted compliances by NCCF and related domains}
	\label{table1}
	\begin{tabular}{lll}
		\hline
		\rowcolor[HTML]{AEAAAA} 
		\textbf{Compliance} & \textbf{Domain} & \textbf{Description}                                                  \\ \hline
		NIST                & Security - IT   & High-level guidelines for   Technology, Security, Innovation          \\
		HIPAA               & Healthcare      & Healthcare data-sensitive   patients’ data and information            \\
		FINRA               & Finance         & Regulations for brokerage   firms, exchange, trade, and finance       \\
		PCI DSS             & Finance         & Regulations for the payment   card industry, security, and data       \\
		COBIT               & Security - IT   & Information technology   management and control process               \\
		Amazon GRC          & IT              & Specific IT compliance   for Amazon developers and customers          \\
		MS GDPR             & IT              & A broad range of   compliances for Microsoft developers and customers \\
		HITECH              & Healthcare      & An act to promote health   information technology                     \\
		CSF                 & Cybersecurity   & Guidelines for   cybersecurity                                        \\ \hline
	\end{tabular}
\end{table}
\section{Milestones, Benchmarks, And Deliverables Roadmap}
	The NCCF strategic roadmap outlines this information technology (IT) enterprise’s current and future capabilities from an infrastructure or systems and technology perspective \cite{jegorova2020information}. This roadmap addresses three major questions when, why, and what solutions should be employed or implemented to enable the enterprise to function and grow in a normal situation and to mitigate potential risks at the lowest possible cost \cite{abbasi2017technology}\cite{park2018role}\cite{shafto2012modeling}.
\subsection{NCCF IT Systems and Technology Roadmap}
The NCCF leadership, during a cross-functional collaboration with IT teams, Human Resource, Legal, and project management teams, created the systems and technology roadmap illustrated in Appendix Figure 4 for three years starting from the first quarter of 2021. The main categories considered in the roadmap included a) infrastructure, b) internal education, c) security and cybersecurity, d) core machine learning development, e) computer vision, f) natural language processing and generation, g) big data, and h) headquarter and geographical expansion. The roadmap assigned each category to an owner team, including a) IT admin, b) Human Resource, c) IT consulting, d) IT security, and e) IT machine learning charged with providing roadmap details for the assigned categories and implement deliverables based on the deadlines. The NCCF roadmap outlined a detailed timeline, including some unchangeable and mandatory deadlines (shown as red blocks). The roadmap’s mandatory deliverables consisted of a) onboarding cloud servers defined in the needs assessment report, b) annual mandatory technical training, c) CV and NLP/NLG team building, and d) a minimal geographical expansion by 2023. 
\newline
The NCCF leadership instructed the owner teams to employ agile project management with quarter-based reporting for flexible milestones and weekly reporting for mandatory deadlines. The roadmap imaged high-level milestones per category, and each assigned team provided the details of milestones, requirements, and deliverables. As shown in Figure \ref{fig1}, the roadmap’s crucial milestones mostly encompassed the cloud-based machine learning services and infrastructure enhancement. The NCCF leadership also aimed to coincide the organization’s infrastructure development to enhance systems’ efficacy and ensure system components’ compatibility. Furthermore, ML-based and analytics services’ development was coincided to decrease the cost-of-service development or onboarding. 
\subsection{Policy and Roadmap Modification Mechanics}
The NCCF leadership instructed the IT admin team to design a cloud-based system to host the NCCF roadmap and policies at the company- and team-level. In this architecture shown in Figure 2, each server and storage (i.e., Amazon S3 bucket) is allocated to each team to create, modify, and store internal policies. The NCCF teams also create the roadmap components and policies requested by the leadership to implement its roadmap. Upon updating or creating new policies, roadmap components, or required entities, the leadership admin server is notified and share the new documents with the leadership for approval. The NCCF IT admin employed a web-based corporate wiki known as Confluence \cite{lindner2011confluence} to share the teams’ policies, company’s policies, and roadmaps with internal staff. 
\newline
The IT admin allocated storage to host Confluence’s data, including the policies, roadmaps, and templates. The internal staff can access Confluence via an internal portal that is attached and controlled by group policies. According to the group policies, a) the company’s roadmap is accessible (visible) by internal staff b) non-managerial staff have only access to their team’s policies and within team roadmaps (no between team roadmaps), c) managerial staff has access to any team’s policies and roadmaps, d) managerial staff can only create, modify, update and deploy their team’s policies and roadmap and e) managerial staff can create, modify and update the company’s roadmap components associated with their team, however, the deployment is conducted after the leadership system admin’s approval. 
\newline
Each team is responsible for team-level policies and roadmap version control hosted in the team’s system; however, the version control of the company’s roadmap projected by Confluence is performed by the leadership system admin, and the previous Confluence versions are stored in the same repository. A backup server is allocated to host a copy of the Confluence storage and has been monthly scheduled to create the backup. Any teams cannot use the Confluence backup server, and they are responsible for creating an internal backup from their documents to ensure the portal’s security.
\begin{figure}[h!]
	\centering
	\includegraphics[width=0.8\linewidth]{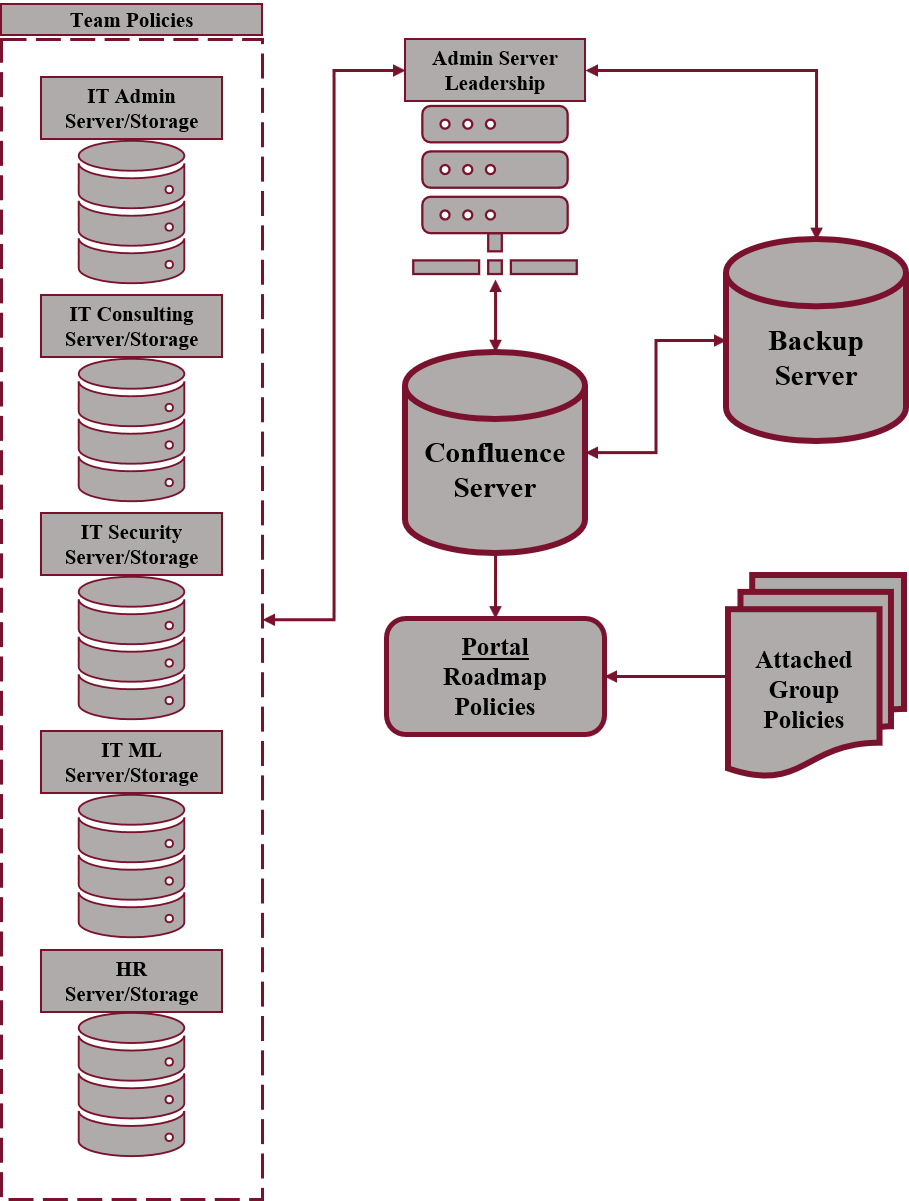}
	\caption{The NCCF roadmap and policies storage and information flow}
	\label{fig2}
\end{figure}
\section{IT Control Framework} 
The NCCF leadership, as a growing organization, designed a control policy framework to ensure software architecture’s quality and functionality. The control framework comprised two central pillars, including a) controls and b) evaluation metrics, which are implemented to control eight major categories of architecture. An automatic system was built to address threats and a reporting module to ensure each major category functions properly by measuring the evaluation metrics and comparing them with predefined policies based on standards. 
\subsection{Network Hardware Devices}
Switches, routers, and other network components form the network hardware devices of the NCCF IT architecture. The NCCF developed the policies and controls for the network and telecommunication inspired by ISO 9001, TL9000, and ESD S20.20 \cite{al2005use}. Network and telecommunication components are controlled by a) security techniques such as encryption, a virtual private network (VPN), b) protection devices such as firewall, proxy server, and the network access server; and c) secure protocols including SSL and SSH.
\subsection{Telecommunications and Network Communication}
The telecommunication and network communication must be secured and encrypted based on Network Communications Standards and Protocols for Ethernet and Wi-Fi based on the 802.11 standards \cite{sun2014wi}. 
\subsection{Operating Systems Access Controls}
The NCCF IT admin provides Linux, macOS, and Windows Server operating systems to its developers, users, and customers. The IT admin team is responsible for upgrading, updating, version controls, and security enhancement.  The operating systems’ functionality (OS), utilization, and log files are monitored using an in-house software application. The group policies defined by the IT admin to control the use of operating systems restrict the staff and customers only to use the operating systems and block any modifications in any form. 
\subsection{Data Control, Metadata and Master Data Management Standards}
The NCCF IT team has established broad data controls, security, and management by adopting and onboarding control policies based on Center of Internet Security (CIS) guidelines \cite{park2018role}\cite{woods2017mapping}. The NCCF IT admin developed 14 categories of controls, including data, metadata, master data management controls, including data quality, audit trail, information security, data custodian, data steward, the data controller, segregation of duties, backup mechanism, data retention, deletion, and disposal, incidents and issues, notifications, and business continuity \cite{macdonald2011pro}\cite{millington2008data}.The data controls defined by NCCF address a variety of data types consisting of abstract, primary, master, personal, event, raw, meta, and reference data \cite{macdonald2011pro}. 
\subsection{Hierarchical Input-Output (HIPO)}
The NCCF leadership developed the HIPO model to provide a visualized version of the hierarchy for system architecture and its modules, including input, process, and output. The HIPO model developed by NCCF to complete customers’ proof of concept projects is illustrated in Appendix Figure 3. The NCCF HIPO model’s general controls include a) leadership approval for each POC step completion and b) project management (PM) scheduling report. 
Application Design and Coding
\newline
The NCCF employs the ISO/IEC 9126 Software engineering Standard for architecture, software design, and implementation (Al-Kilidar et al., 2005). The general controls enforced include software manager and project lead approval before release deliverables passing through standard software testing processes. The software codes must be passed through an automatic code quality review tool referring to Black as a Git pre-commit\footnote{https://ljvmiranda921.github.io/notebook/2018/06/21/precommits-using-black-and-flake8/} application. 
\subsection{Machine Learning Services and Applications}
The central pillar of machine learning applications is data provided by customers or public sources; therefore, data protection and security, and different compliances play crucial roles \cite{sachs2009toward}. Also, machine learning modeling democratization refers to avoiding intentional bias in sensitive topics such as gender and race during data collection and model development \cite{kross2020democratization}. The same data proception controls previously discussed are enforced for the machine learning category. Also, POC project leads collaborating with a customer to define a list of evaluation metrics depending on the nature of POC. 
\subsection{Business Intelligence (BI) Tools}
\begin{figure}[h!]
	\centering
	\includegraphics[width=0.8\linewidth]{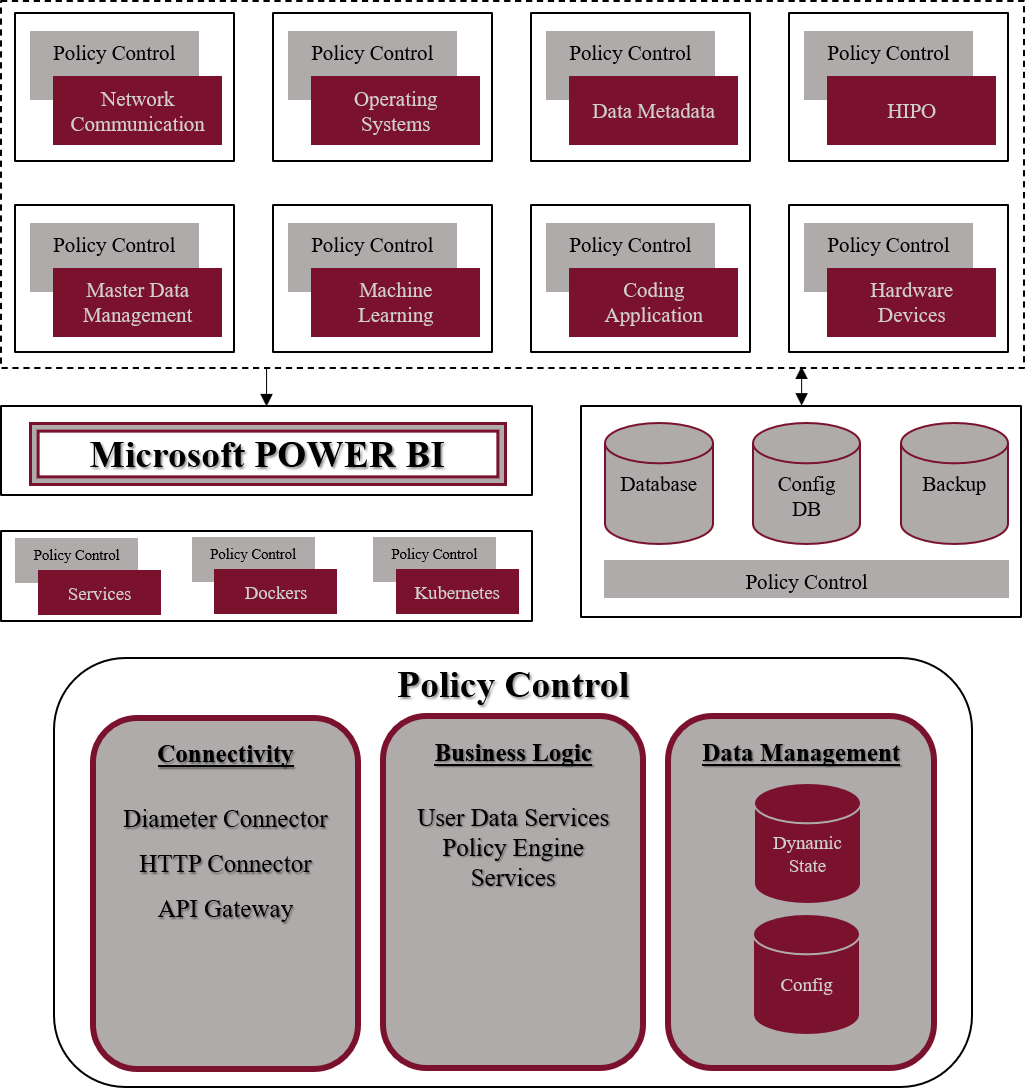}
	\caption{The control framework implemented by NCCF (top), policy control components (bottom)}
	\label{fig3}
\end{figure}

T
The NCCF has recommended that the PM team employs a commercial business intelligence tool known as Microsoft Power BI\footnote{https://powerbi.microsoft.com/en-us/what-is-power-bi/}  to speed up dashboard implementation. Microsoft Power BI offers numerous capabilities, including a) data connection (Excel, CSV, SQL, Azure), b) data relationships from single or multiple data modeling, c) power query and pivot access, d) custom and enhanced visualization, e) report sharing and f) mobile applications. The NCCF leadership also requested the PM team to explore other commercial BI tools such as SAP Business Intelligence and Tableau as a backup plan. The NCCF policy control framework is illustrated in Figure \ref{fig3}, including the eight main components, Microsoft POWER BI as the reporting system, and database units along with the components of policy control blocks.
\section{Stakeholders Responsibilities}
\begin{figure}[h!]
	\centering
	\includegraphics[width=0.8\linewidth]{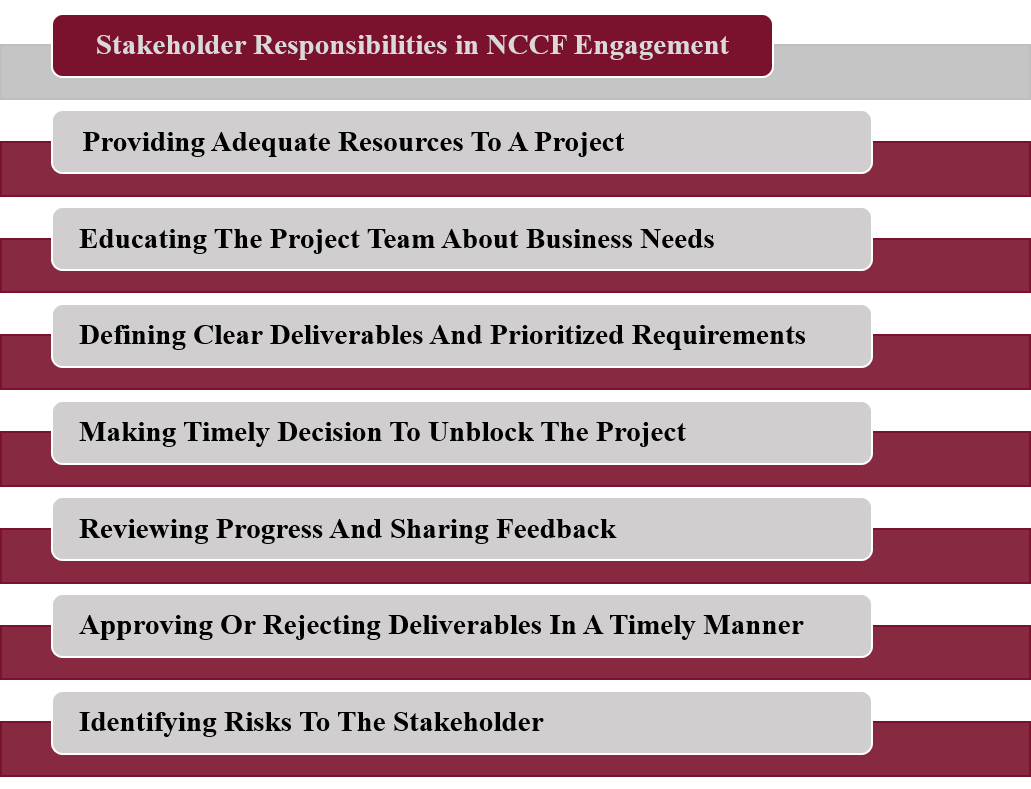}
	\caption{Stakeholder responsibilities in NCCF engagements}
	\label{fig4}
\end{figure}

NCCF considers engagements with its customers as collaborative projects where both parties have shared and separated responsibilities. The NCCF leadership and legal team have implemented a statement of work template indicating customers’ responsibilities. Customers’ primary responsibility refers to data ownership, security, and protection \cite{sachs2009toward}. Although NCCF employs all possible mechanisms to protect customers’ data, the customers are responsible for data protection in their accounts unless they use a purchased service from NCCF or its cloud providers \cite{fassin2012stakeholder}. Figure \ref{fig4} illustrates multi-step stakeholders’ responsibilities defined by the NCCF leadership.  
\section{Communication and Training Plan}
\begin{figure}[h!]
	\centering
	\includegraphics[width=0.5\linewidth]{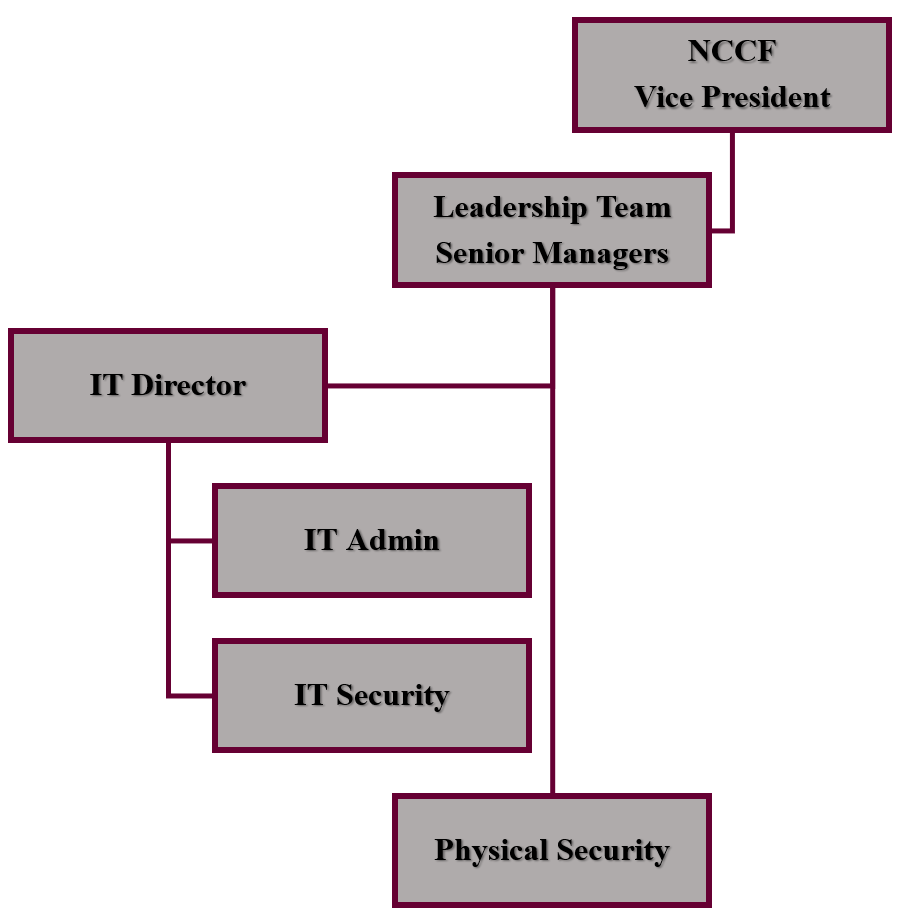}
	\caption{The control framework implemented by NCCF (top), policy control components (bottom)}
	\label{fig5}
\end{figure}

The NCCF knowledge transfer plan (KTP) covers 15 major threats to the organization. It consists of a) service category, b) risk level, c) service layer, d) key information, e) trainees, f) training activities, g) scheduling, and h) documents location illustrated in Table \ref{table2}. The architecture of NCCF KTP included a) Confluence user interface as the knowledge sharing tool, b) five system layers, and c) the database hosting the data and materials. The AWS cloud environment hosts the NCCF KTP architecture, and a mirror server hosts the backup of KTP data. Furthermore, to address each threat, the NCCF PRM recommends relocating some activities and providing instructions for business continuity discussed in the contingency plan \cite{garson2006public}\cite{kassema2016disaster}.
\newline
The NCCF PRM implemented an emergency communications plan to provide guidelines and instructions to staff, including IT teams, to show immediate actions in unexpected incidents and emergencies \cite{shafto2012modeling} . The company’s emergency plan was designed based on eight factors, including a) launching fast, b) briefing leadership, c) identifying spokesperson, d) preparing company’s statement for customers’ briefing, e) triggering IT backup services f) switching networks to a backup service provider g) relocating staff in danger in a safe place and h) estimating time of recovery and continuously updating \cite{greer2012personal}. The NCCF emergency plan included the instructions for three major IT categories a) data, b) network, and c) services and physical safety and emergency shown in Table 3 in Appendix. The hierarchy of reporting in the emergency plan includes direct communication between the NCCF vice president, the leadership team, and IT managers shown in Figure \ref{fig5}.

\section{Leadership Involvement}
The NCCF leadership is ultimately responsible for growth, failure, IT or physical incidents and risks that could potentially occur. Therefore, in the company’s IT policy, control, contingency (Table 4), and emergency plan, its leadership is the gateway of communication and decision-making (Maringe, 2012). The leadership involvement in NCCF engagements with customers or incidents enables the company’s staff and developers to sustain decisions and to mitigate risks to the organization. Although the company’s IT risk plan provides details of risk mitigation processes, the leadership involvement in final decision-making during mitigating risks expedites the business recovery (Maringe, 2012). Also, many customers have internal metrics to evaluate vendors, and the research showed that the leadership involvement demonstrates the customer obsession with vendors, including NCCF (Ejimabo, 2015).
\clearpage
\section{Conclusion}
The NCCF IT policy plan discussed in this work consisted of several components such as a) proposed policies, b) milestones, deliverables, and roadmaps, c) control framework, d) customers’ responsibilities, and e) emergency communication plan and leadership involvement. The data privacy, security, and business continuity were considered central pillars of the NCCF IT policy plan as major NCCF’s customers seek data migration and data-driven services. The NCCF leadership provided several evaluation metrics to measure the plan’s performance; however, the plan could be updated upon internal or external feedback if required. The NCCF leadership noticed that the proposed plan highly concentrated on the data categories. Therefore, the company’s leadership should augment the IT policy plan’s content by equalizing the importance of other aspects such as services (machine learning solutions) and infrastructure soon.
\clearpage
\bibliographystyle{ieeetran}
\bibliography{references}  





\appendix
\clearpage
\begin{table}[h!]
	\caption{The NCCF KTP (primary components)}
	\label{table2}
	\resizebox{\textwidth}{!}{%
		\begin{tabular}{llllll}
			\hline
			\rowcolor[HTML]{AEAAAA} 
			No. & Knowledge                                 & Category & Risk Level   & Service Layer                                                                                                                              & Key Information                                                                                                 \\ \hline
			\rowcolor[HTML]{FFFFFF} 
			1   & Abuse of Cloud   Services                 & Service  & Minor        & \begin{tabular}[c]{@{}l@{}}1 - Escalates to Service Layer 2\\ 2 - Electronic Support \\ 4 - External Service Providers\end{tabular}        & \begin{tabular}[c]{@{}l@{}}Group Policy\\ Legal Aspects\\ Monitoring\end{tabular}                               \\
			\rowcolor[HTML]{FFFFFF} 
			2   & Account or Service Hijacking              & Service  & Catastrophic & \begin{tabular}[c]{@{}l@{}}1 - Escalates to Service Layer 2\\ 2 - Electronic Support \\ 5 - Activation of Contingency Plans\end{tabular}   & \begin{tabular}[c]{@{}l@{}}Phishing\\ Anti-Virus\\ Firewall\\ Personal Security\\ Network Security\end{tabular} \\
			\rowcolor[HTML]{FFFFFF} 
			3   & Advanced Persistent   Threats (APTs)      & Network  & Significant  & \begin{tabular}[c]{@{}l@{}}1 - Escalates to Service Layer 2\\ 2 - Electronic Support \\ 5 - Activation of Contingency Plans\end{tabular}   & \begin{tabular}[c]{@{}l@{}}Cyberattacks\\ Defense Mechanisms\end{tabular}                                       \\
			\rowcolor[HTML]{FFFFFF} 
			4   & Communication with Cloud Service Provider & Network  & Moderate     & \begin{tabular}[c]{@{}l@{}}3 - On-Site Service Support\\ 4 - External Service Providers\end{tabular}                                       & \begin{tabular}[c]{@{}l@{}}Cloud Security\\ Network Architecture\end{tabular}                                   \\
			\rowcolor[HTML]{FFFFFF} 
			5   & Data Breaches                             & Data     & Moderate     & 5 - Activation of Contingency Plans                                                                                                        & \begin{tabular}[c]{@{}l@{}}Stolen   Information\\ Password Guessing\\ Phishing\\ Prevention\end{tabular}        \\
			\rowcolor[HTML]{FFFFFF} 
			6   & Data Loss                                 & Data     & Major        & \begin{tabular}[c]{@{}l@{}}1 - Escalates to Service Layer 2\\ 2 - Electronic Support\end{tabular}                                          & \begin{tabular}[c]{@{}l@{}}Insufficient Expertise\\ Recklessness\end{tabular}                                   \\
			\rowcolor[HTML]{FFFFFF} 
			7   & Distributed   Denial of Services (DDoS)   & Network  & Catastrophic & \begin{tabular}[c]{@{}l@{}}1 - Escalates   to Service Layer 2\\ 2 - Electronic Support \\ 5 - Activation of Contingency Plans\end{tabular} & \begin{tabular}[c]{@{}l@{}}Cyberattacks\\ Defense Mechanisms\end{tabular}                                       \\
			\rowcolor[HTML]{FFFFFF} 
			8   & Inadequate Access Management              & Data     & Significant  & \begin{tabular}[c]{@{}l@{}}1 - Escalates to Service Layer 2\\ 2 - Electronic Support\end{tabular}                                          & \begin{tabular}[c]{@{}l@{}}Group Policy\\ Proper Roles Allocation\end{tabular}                                  \\
			\rowcolor[HTML]{FFFFFF} 
			9   & Insecure APIs                             & Data     & Catastrophic & 5 - Activation   of Contingency Plans                                                                                                      & \begin{tabular}[c]{@{}l@{}}Penetration Testing\\ Security\\ Network Traffic\end{tabular}                        \\
			\rowcolor[HTML]{FFFFFF} 
			10  & Insufficient Due Diligence                & Data     & Major        & \begin{tabular}[c]{@{}l@{}}3 - On-Site Service Support\\ 4 - External Service Providers\end{tabular}                                       & Legal Aspects                                                                                                   \\
			\rowcolor[HTML]{FFFFFF} 
			11  & Malicious   Insider Threats               & Service  & Minor        & 3 - On-Site   Service Support                                                                                                              & \begin{tabular}[c]{@{}l@{}}Violence\\    Spy\end{tabular}                                                       \\
			\rowcolor[HTML]{FFFFFF} 
			12  & Malware Injection                         & Service  & Significant  & 2 - Electronic Support                                                                                                                     & \begin{tabular}[c]{@{}l@{}}Malware\\ Anti-Virus\\ Firewall\\ Personal Security\end{tabular}                     \\
			\rowcolor[HTML]{FFFFFF} 
			13  & Poor Internet   Protocol Protection       & Network  & Minor        & \begin{tabular}[c]{@{}l@{}}3 - On-Site   Service Support\\ 4 - External Service Providers\end{tabular}                                     & \begin{tabular}[c]{@{}l@{}}Network Configuration\\ Network Security\end{tabular}                                \\
			\rowcolor[HTML]{FFFFFF} 
			14  & Shared Technology                         & Network  & Major        & \begin{tabular}[c]{@{}l@{}}2 - Electronic Support \\ 3 - On-Site Service Support\end{tabular}                                              & \begin{tabular}[c]{@{}l@{}}Cloud Architecture\\ Group Policy\end{tabular}                                       \\
			\rowcolor[HTML]{FFFFFF} 
			15  & System   Vulnerabilities                  & Service  & Major        & \begin{tabular}[c]{@{}l@{}}4 - External   Service Providers\\ 5 - Activation of Contingency Plans\end{tabular}                             & \begin{tabular}[c]{@{}l@{}}Security\\ System Configurations\end{tabular}                                        \\ \hline
		\end{tabular}%
	}
\end{table}
\begin{table}[h!]
	\caption{Emergency Communications Plan with detailed instructions}
	\label{table3}
	\resizebox{\textwidth}{!}{%
		\begin{tabular}{llll}
			\hline
			\rowcolor[HTML]{9B9B9B} 
			\textbf{Category} & \textbf{Action}                                                                                                                                                                                                                                                                                             & \textbf{Staff in Charge} & \textbf{Contact} \\ \hline
			Data              & \begin{tabular}[c]{@{}l@{}}In case of database and server disruption for \\ more than 4 hours, the engaged team contacts \\ the service provider to spin up back servers and \\ temporary cloud instances to address customers’ needs.\end{tabular}                                                         & IT director              & 650-000-0000     \\
			Network           & \begin{tabular}[c]{@{}l@{}}In case of network disruption, immediately \\ contact the main service provider to switch \\ the network to another service. In case of a major issue \\ with the main provider, immediately contact the \\ backup network provider to switch networks temporarily.\end{tabular} & IT Admin Manager         & 408-000-0000     \\
			Services          & \begin{tabular}[c]{@{}l@{}}Immediately contact the service team and account manager\\  at AWS to spin up temporary servers and redirect customers’ \\ and company’s traffic to the temp servers\end{tabular}                                                                                                & IT Admin Manager         & 408-000-0000     \\
			Physical Safety   & \begin{tabular}[c]{@{}l@{}}Immediately contact the physical security team \\ in case of violence. In case of natural incidents \\ such as fire and earthquake, follow the security \\ instructions and evacuate the facilities.\end{tabular}                                                                & Security                 & 628-000-0000     \\ \hline
		\end{tabular}%
	}
\end{table}
\clearpage
\begin{figure}[ht!]
	\centering
	\includegraphics[width=0.8\linewidth]{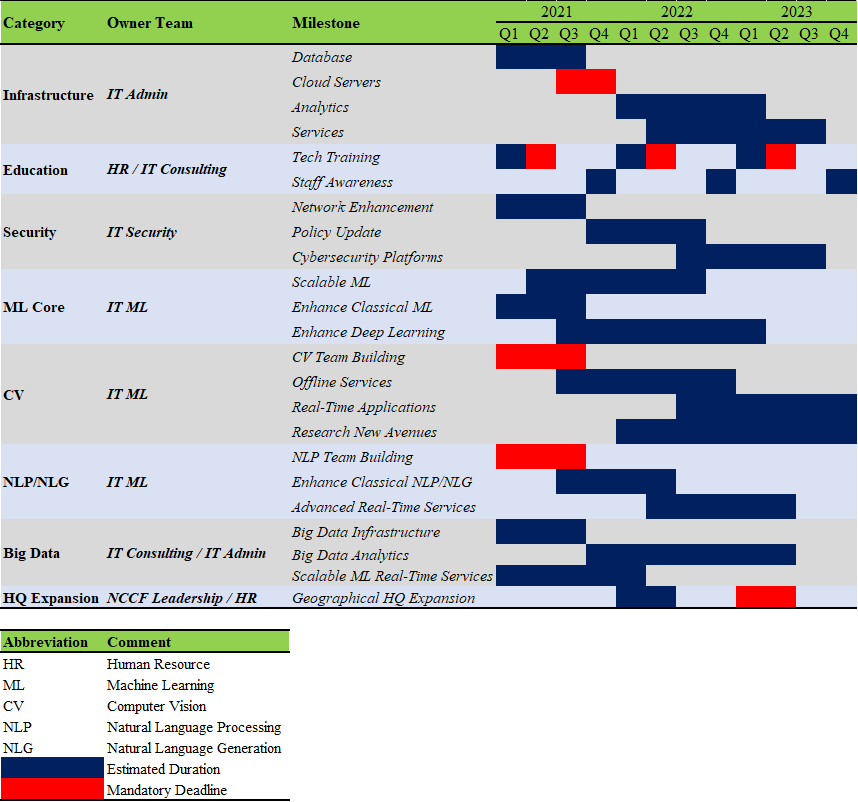}
	\caption{The NCCF systems and technical roadmap}
	\label{fig6}
\end{figure}
\end{document}